\documentclass[aps,twocolumn,pra,superscriptaddress,showpacs,tightenlines]{revtex4}
\usepackage{amssymb}
\usepackage{amsmath}
\usepackage{graphicx}
\usepackage{epsfig}
\usepackage{subfigure}
\usepackage{amsfonts}

\begin{document}

\title{Controlling soliton excitations in Heisenberg spin chain through
magic angle}
\author{Jing Lu}
\affiliation{Department of Physics, Hunan Normal University, Changsha 410081, China}
\affiliation{Institute of Theoretical Physics, Chinese Academy of Sciences, Beijing,
100080,China}
\author{Lan Zhou}
\affiliation{Department of Physics, Hunan Normal University, Changsha 410081, China}
\author{Le-Man Kuang}
\affiliation{Department of Physics, Hunan Normal University, Changsha 410081, China}
\author{C. P. Sun}
\email{suncp@itp.ac.cn} \homepage{http://www.itp.ac.cn/~suncp}
\affiliation{Institute of Theoretical Physics, Chinese Academy of
Sciences, Beijing, 100080,China}

\begin{abstract}
We study the nonlinear dynamics of collective excitation in a $N$-site $XXZ$
quantum spin chain, which is manipulated by an oblique magnetic field. We
show that, when the tilted field is applied along the magic angle $%
\theta_{0} =\pm\arccos \sqrt{1/3}$, the anisotropic Heisenberg spin
chain becomes isotropic and thus an free propagating spin wave is
stimulated. And in the regime of the tilted angle larger and smaller
then the magic angle, two types of nonlinear excitations appear,
which are bright soliton and dark soliton.
\end{abstract}

\maketitle

\section{\label{Sec:introduction}Introduction}

A crucial challenge for quantum information processing (QIP) is perfectly
transmitting a quantum state from one place to another as well as storing
the information of a quantum state. Recently, as much attentions have been
paid to the solid-based quantum computing, there are many proposals for
quantum state transfer and storage, using quantum spin systems as quantum
data bus and quantum memory\cite{BosePRL91,sunPRA71-1,sunPRA71-2,chrisPRL92}.

In other hand the Heisenberg chain is a typical spin system in
condensed-matter physics. It has attracted considerable attentions for a
long time in describing various magnetic properties of materials. As a
strong correlated system\cite%
{Faddeev-87,Kivshar-RMP8991,Baryakhtar-94,Huang-PRl95,Fogler-PRL02}, it can
display rich nonlinearities induced by inter-spin interaction. The typical
nonlinear phenomenon is the solitary wave, that is a wave packet propagate
without either energy loss and wave packet spreading. In this sense it is
quite natural to consider the role of soliton wave in quantum state transfer
and quantum information storage.

Generally there are two major approaches to theoretically study the solitary
excitation in the one-dimension (1D) Heisenberg chain\cite%
{Kosevich-PhR90,JTjon-PRB77,Fogedby-JPA80,Pushkarov-PLA77,
Balakrishnan-PRL85,RFerrer-PRB89,Skrinjar-JPh89,GHuang-PRB90,Daniel-PRA99}.
1) each spin is represented as a Bloch vector, and a solitary excitation is
found by investigating the nonlinear dynamics of this classical variable\cite%
{JTjon-PRB77,Fogedby-JPA80}. 2) Another procedure is to employ the boson
mappings of spin operators via the Holstein-Primakoff transformation and
Jordan-Schwinger realization\cite{Slichter-96}. By making use of the
spin-coherent-state representation, a solvable nonlinear differential
equation can be deduced from the Heisenberg equation for the mapping boson
operators, therefore a solitary excitation is implied in the system\cite%
{Pushkarov-PLA77,Balakrishnan-PRL85,RFerrer-PRB89,Skrinjar-JPh89,GHuang-PRB90,Daniel-PRA99}%
.

Our present investigation is motivated by the well known knowledge in
nuclear magnetic resonance (NMR) that the effective spin-spin interaction
can be controlled by the angle of the external applied magnetic field\cite%
{Slichter-96}. Here, to some extent, the inter-spin coupling can be canceled
effectively by the magic angle . Nowadays together with the spin echo
technology, magic angle manipulation has became a necessary tool for
controlling the interacting spin system to be free of decoherence due to the
inter-spin couplings. In this paper, we apply a magnetic field on a N-site
XXZ spin chain, which is rotated around the y-axis by an amount of magnitude
$\theta$, and study how to manipulate a soliton excitation by this oblique
field. It is known that the ferromagnetic order is completely determined by
the direction of the external magnetic field. Indeed, our investigation
shows that the switch between the bright soliton and the dark soliton is
controlled by $\theta$, whose tangent is defined as the ratio between the z
component and the x component of the magnetic field; and at the turning
magnitude referred to as the magic angle $\theta_{0}$, the original
anistropic chain becomes isotropic, therefore only an ideal spin wave is
stimulated as an effective plane wave. Here, we employ the second approach
mentioned above, but go beyond the spin wave approach by considering the
nonlinear effect of the collective spin excitation. Generally, the spin wave
approach is regarded as a mean field method with a given order parameter.
Due to the ignorance of the nonlinear fluctuation, the spin wave approach
does not cover the quantum fluctuation in the nonlinear regime. However our
approach overcomes this disadvantage.

The remainder of this paper is organized as follows. In section~\ref
{Sec:model}, we present our model -- a Heisenberg spin chain with
anisotropic coupling in an oblique magnetic field. The approach we
use is in sections~\ref{Sec:quasi-c} and~\ref{Sec:c-fild}. Here the
quasi-classical equation of motion is obtained, which is a nonlinear
Schr\"{o}dinger (NLS) equation. In section~\ref{Sec:time-e}, the
time evolution of nonlinear excitations is studied. The bright
soliton and dark soliton can be excited
as $\theta$ shifted to the left or the right of the magic angle. In section~%
\ref{Sec:conclusion}, we give a remark about the relationship between the
soliton wave propagation and perfect quantum state transfer.

\section{\label{Sec:model}A Spin Chain Model}

We consider a one-dimensional XXZ spin chain with $N$ spins in an
homogeneous external magnetic field
\begin{equation}
\mathbf{B}=B\left( e_{x}\sin \theta +e_{z}\cos \theta \right)  \label{2-02}
\end{equation}
Denote the spin operator at the $j$th site by Pauli spin operator $S_{j}^{x}$%
, $S_{j}^{y}$, and $S_{j}^{z}$. Then Hamiltonian $H$ of this system reads
\begin{equation}
H=-J\sum_{j=1}^{N}\left( \hat{S}_{j}^{x}\hat{S}_{j+1}^{x}+\hat{S}_{j}^{y}%
\hat{S}_{j+1}^{y}+\Delta \hat{S}_{j}^{z}\hat{S}_{j+1}^{z}\right) +\mathbf{B}%
\cdot \mathbf{S}  \label{2-01}
\end{equation}
where $\mathbf{S=}\sum_{j}\mathbf{S}_{j}$ is the total spin operator. Three
types of interactions are included in Eq.(\ref{2-01}): 1) the isotropic part
of nearest neighbor exchange interaction $J_{x}=J_{y}=J>0$. 2) the
anisotropic part of nearest neighbor exchange interaction, which is
characterized by a dimensionless parameter $\Delta =J_{z}/J$. 3) the
external homogeneous field $\mathbf{B}$, which is specified by its
components $B_{x}=B\sin \theta $ and $B_{z}=B\cos \theta $. Here the
magnitude of the magnetic field $B=\sqrt{ B_{x}^{2}+B_{z}^{2}}$, and the
oblique angle $\theta =\arctan (B_{z}/B_{x})$.

Obviously, the last term is proportional to the total spin $z$-component $%
\hat{S}_{\mathrm{total}}^{z}=\sum_{\mathbf{i}}\hat{S}_{\mathbf{i}}^{z}$,
which is conserved and thus has common eigenstates with the total
Hamiltonian. Correspondingly, the Hilbert space of the system can be
decomposed into a direct sum of numerous subspaces $V(M)$ specified by the
total spin number $M$ along z-axis. It had been proved~\cite{Lieb,Affleck}
that, on a finite simple cubic lattice, the ground state of the $XXZ$ model
is nondegenerate in an subspace $V(M)$. In particular, its global ground
state $\Psi _{0}(\>\Delta )$ is just the ground state of the model in the
subspace $V(M=0)$ When the lattice is finite, the ground state energy $%
E_{0}(\Delta )$ and the spin correlation function $\langle \hat{S}_{\mathbf{i%
}}^{z}\hat{S}_{\mathbf{i}}^{z}\rangle $ are analytical with respect to $%
\Delta $.

Here, we assume that the external magnetic field is much larger than the
inter-spin interaction. Therefore we switch to a new ``reference frame''
with new $Z-$axis along the quantized direction of spin along the effective
field $\mathbf{B}$. The corresponding $\mathbf{S}_{j}\rightarrow \mathbf{L}%
_{j}$ transformation for each spin at site $j$ reads
\begin{align}
L_{j}^{x}& =\hat{S}_{\mathbf{j}}^{x}\cos \theta -\hat{S}_{\mathbf{j}%
}^{z}\sin \theta ,  \notag \\
L_{j}^{y}& =\hat{S}_{\mathbf{j}}^{y}, \\
L_{j}^{z}& =\hat{S}_{j}^{z}\cos \theta +\hat{S}_{j}^{x}\sin \theta .  \notag
\end{align}%
Therefore Hamiltonian in Eq.~(\ref{2-01}) is decomposed into a direct sum of
the $SO(3)$ irreducible tensors with respect to $\mathbf{L}_{j}$ , i.e,
\begin{equation}
H=BL_{z}+\sum_{M=-2}^{M=2}H_{M}  \label{h-comp}
\end{equation}%
which is written according to the irreducible representation $D^{[L]}$ $%
(L=2,1,0)$of $SO(3)$ group. Here, $L_{z}=\sum_{j}L_{j}^{z}$ is the third
component of the total angular momentum. With $\delta =\left( \Delta
-1\right)$,
\begin{eqnarray}
H_{0} &=&-\left( 1+\frac{\delta }{2}\sin ^{2}\theta \right) J\sum_{j}\mathbf{%
L}_{j}\cdot \mathbf{L}_{j+1}  \notag \\
&&-\frac{\delta }{2}J\left( 3\cos ^{2}\theta -1\right)
\sum_{j}L_{j}^{z}L_{j+1}^{z}
\end{eqnarray}%
belongs to the zero rank representation $D^{[0]}$ of $SO(3)$, while
\begin{equation}
H_{\pm 1}=\frac{\delta }{4}J\sin 2\theta \sum_{j}\left(
L_{j}^{z}L_{j+1}^{\pm }+L_{j}^{\pm }L_{j+1}^{z}\right)
\end{equation}%
and
\begin{equation}
H_{\pm 2}=-\frac{\delta }{4}J\sin ^{2}\theta \sum_{j}L_{j}^{\pm
}L_{j+1}^{\pm }
\end{equation}%
belong to the 1st and 2nd rank representation $D^{[1]}$ and $D^{[2]}$
respectively. The $M$th order tensors $H_{M}$ satisfy
\begin{equation}
\lbrack L_{z},H_{M}]=MH_{M},[L_{z},H_{0}]=0.  \label{so3}
\end{equation}

From the point of view of perturbation theory, the eigenvalues of
the Hamiltonian are those of $B L_{z}$ at the $0$th order, i.e.
$L_{z}=M$, which means the ground state has all spins parallel to
the field and it is not degenerate. However, the excited state,
where one or two spins are flipped with respect to the ground state,
are degenerate. Therefore, when the perturbation theory is applied,
we diagonalize Hamiltonian $H$ in each subspace of a given $M$,
which means that at lowest order, Hamiltonian $H_{0}$ is kept. Thus,
the $H_{0}$ plays an indispensable role to govern the dynamics of
the spin chain. It can be found directly from Hamiltonian $H_{0}$
that the spin-spin interaction is controlled by the direction of the
applied magnetic field. When $\theta=\theta_{0}=\pm\arccos
\sqrt{1/3}$, the original anisotropic chain becomes isotropic.

We further explain the above argument from the point view of the
representation theory of SO(3) together with rotating wave
approximation. Sandwiched by the common states $ |J,M^{\prime
}\rangle $ and $|J,M^{\prime \prime }\rangle $ of $H_{0}$ and $
L_{z}$, the equations (\ref{so3}) lead to
\begin{eqnarray}
&&\langle J,M^{\prime }|[L_{z},H_{M}]|J,M^{\prime \prime }\rangle  \notag \\
&=&(M^{\prime }-M^{\prime \prime })\langle J,M^{\prime }|H_{M}|J,M^{\prime
\prime }\rangle \\
&=&M\langle J,M^{\prime }|H_{M}|J,M^{\prime \prime }\rangle  \notag
\end{eqnarray}%
here $M=M^{\prime }-M^{\prime \prime }$ for the nonvanishing matrix
element $\langle J,M^{\prime }|H_{M}|J,M^{\prime \prime }\rangle$.
In the interaction picture, the off-diagonal elements $\langle
J,M^{\prime }|H_{M}|J,M^{\prime \prime }\rangle $ are fast changing
with high frequency, therefore the terms with $|M|=1$ and $|M|=2$ in
the total Hamiltonian can be ignored~\cite{Slichter-96}.

\section{\label{Sec:quasi-c}quasi-classical motion equation for
ferromagnetic spin chain}

With the above considerations, the $H_{M}$ $(M=\pm 1,\pm 2)$ are the first
and second order tensor operators that transform according to the
representation $D^{[j=1,2]}$ of $SO(3)$ group. In large external field
limit, Hamiltonian (\ref{h-comp}) is reduced to
\begin{equation}
H=BL_{z}+H_{0}.  \label{h-simp}
\end{equation}%
In this section, we begin with the above Hamiltonian to discuss the soliton
excitations.

To describe the spin excitation, we introduce the boson excitation by the
Holstein-Primakoff transformation~\cite{HP-PR40},
\begin{subequations}
\label{hp}
\begin{eqnarray}
L _{i}^{+}& =\hat{a}_{j}^{\dag }\sqrt{2S-\hat{a}_{j}^{\dag }\hat{a}_{j}},
\label{hp-s+} \\
L _{i}^{-}& =\sqrt{2S-\hat{a}_{j}^{\dag }\hat{a}_{j}}\hat{a}_{j},
\label{hp-s-} \\
L _{i}^{z}& =\hat{a}_{j}^{\dag }\hat{a}_{j}-S,  \label{hp-sz}
\end{eqnarray}
where the annihilation operators $\hat{a}_{i}$ and the creation operators $%
\hat{a}_{j}^{\dag }$ satisfy the boson commutation relation $\left[ \hat{a}%
_{i},\hat{a}_{j}^{\dag }\right]=\delta _{ij}$. The number operator $\hat{n}%
_{i}=\hat{a}_{i}^{\dag }\hat{a}_{i}$ characterizes the spin deviation from
its maximum value $S$ of $L_{i}^{z}$. Usually, for a Heisenberg
ferromagnetic system with inter-spin couplings $-J\mathbf{L}_{i}\cdot
\mathbf{L}_{i+1}$ ($J>0)$, the spontaneous symmetry breaking will happen
along the direction of the external magnetic field (says along $z$-axis),
which is subsequently assumed to approach zero. The ground state implies a
ferromagnetic order with all spins along the $+z$ or $-z$ direction.
Therefore the magnetization comes into being, which is defined as the
non-vanishing average of $L^{z}=\sum_{i}L_{i}^{z}$. The governing equation
for the nonlinear excitations on ground states can be obtained by the
expanding $\sqrt{2S-\hat{n}_{i}}\simeq \sqrt{2S}(1-\hat{n}_{i}/(4S)) $ to a
series over the low excitation $\langle \hat{n}_{i}\rangle \ll 2S$. We first
expand (\ref{hp-s+}) and (\ref{hp-s-}), and keep the terms in the first
order of $\langle\hat{n}_{i}\rangle$
\end{subequations}
\begin{subequations}
\begin{eqnarray}
L _{i}^{+}& \approx \sqrt{2S}\hat{a}_{j}^{\dag }\left( 1-\frac{1}{4S}\hat{a}%
_{j}^{\dag }\hat{a}_{j}\right) ,  \label{hp-s+app} \\
L _{i}^{-}& \approx \sqrt{2S}\left( 1-\frac{1}{4S}\hat{a}_{j}^{\dag }\hat{a}%
_{j}\right) \hat{a}_{j}.  \label{hp-s-app}
\end{eqnarray}

Consequently, beyond the spin wave approximation, the low energy effective
Hamiltonian (\ref{h-simp}) is achieved as
\end{subequations}
\begin{eqnarray}
H &=&-c_{0}S\sum_{j}\left( \hat{a}_{j}^{\dag }\hat{a}_{j+1}+h.c\right)
+B\sum_{j}n_{j}  \notag \\
&&+\frac{c_{0}}{4}\sum_{j}\left( \hat{a}_{j}^{\dag }n_{j}\hat{a}_{j+1}+n_{j}%
\hat{a}_{j}\hat{a}_{j+1}^{\dag }+h.c\right)  \notag \\
&&-\left( c_{0}+c_{1}\right) \sum_{j}n_{j}\left(n_{j+1}-2S\right) ,
\label{h-hp}
\end{eqnarray}
where $n_{j}=\hat{a}_{j}^{\dag }\hat{a}_{j}$, $c_{0}=J\left( 1+\delta \sin
^{2}\theta /2\right) $ and $c_{1}=\delta J\left( 3\cos ^{2}\theta -1\right)
/2$.

As we should emphasize that, the external magnetic field has been assumed $%
B>0$ in the above discussions. If $B<0$, another form of Holstein-Primakoff
transformation should be taken
\begin{subequations}
\begin{eqnarray}
L _{i}^{+}& =\sqrt{2S-\hat{a}_{j}^{\dag }\hat{a}_{j}}\hat{a}_{j}, \\
L _{i}^{-}& =\hat{a}_{j}^{\dag }\sqrt{2S-\hat{a}_{j}^{\dag }\hat{a}_{j}}, \\
L _{i}^{z}& =S-\hat{a}_{j}^{\dag }\hat{a}_{j},
\end{eqnarray}
but the main result of this paper is same. So without loss of generality, $%
B>0$ is assumed in the rest of this paper.

Hamiltonian (\ref{h-hp}) characterizes the low energy nonlinear property of
ferromagnetic spin chain in an oblique magnetic field. The corresponding
Heisenberg equation
\end{subequations}
\begin{eqnarray}
i\hbar \frac{d\hat{a}_{j}}{dt} &=&-c_{0}S\left( \hat{a}_{j+1}+\hat{a}%
_{j-1}\right) +\left( 2S\left( c_{0}+c_{1}\right) +B\right) \hat{a}_{j}
\notag \\
&&+\frac{c_{0}}{4}\left( 2n_{j}\hat{a}_{j\pm1}+n_{j\pm1}\hat{a}_{j\pm1} +%
\hat{a}_{j\pm1}^{\dag }\hat{a}_{j}^{2}\right)  \notag \\
&&-\left( c_{0}+c_{1}\right) \hat{a}_{j}n_{j\pm1}
\label{operator-Heisenbergeq}
\end{eqnarray}
contains various nonlinear couplings. They lead to different nonlinear
``phases'', which are represented by various types of soliton excitations.

\section{\label{Sec:c-fild}Continuum field approach for nonlinear collective
excitation}

In this section, we make use of the continuum field theory to study the
nonlinear excitations in the ferromagnetic spin chain. To this end we
consider the spin wave approach with some nonlinear corrections. First let
us introduce the $p-$representation (or call the Glauber coherent state
representation \cite{Glauber-PR63}) defined by the product of the multi-mode
coherent states $\left\vert \alpha \right\rangle =\prod_{i}\left\vert \alpha
_{i}\right\rangle $, where each component $\left\vert \alpha
_{i}\right\rangle$ is the eigenstate of the annihilation operator $\hat{a}%
_{i}$, i.e., $\hat{a}_{i}\left\vert \alpha _{i}\right\rangle =\alpha
_{i}\left\vert \alpha _{i}\right\rangle$, and $\alpha_{i}$ is the coherent
amplitude. Since coherent states are normalized and overcompleted, the field
operator sandwiched by $\left\vert \alpha \right\rangle$ can be represented
only with their diagonal elements. Thus, we only need to consider the
diagonal part of Eq. (\ref{operator-Heisenbergeq}), which are enough to
describe the nonlinear dynamics without any help of off-diagonal elements.
The p-representation of non-linear equations (\ref{operator-Heisenbergeq})
reads 
\begin{eqnarray}
i\hbar \frac{d\alpha _{j}}{dt} &=&-c_{0}S\left( \alpha _{j\pm1} -2\alpha
_{j}\right) +\left( 2Sc_{1}+B\right) \alpha _{j}  \notag \\
&&+\frac{c_{0}}{4}\left( 2\left\vert \alpha _{j}\right\vert ^{2}\alpha
_{j\pm1}+\left\vert \alpha _{j\pm1}\right\vert ^{2}\alpha _{j\pm1}+\alpha
_{j\pm1}^{\ast }\alpha _{j}\alpha _{j}\right)  \notag \\
&& -\left(c_{0}+c_{1}\right) \alpha _{j} \left\vert \alpha
_{j\pm1}\right\vert^{2}.  \label{coherent amplitude-Heisenbergeq}
\end{eqnarray}

From then on, the spin dynamics is expressed in terms of the $c-$number
equation in Glauber's coherent-state representation. However Eq.~(\ref%
{coherent amplitude-Heisenbergeq}) is difficult to solve due to its
nonlinearity and discreteness. Since we are looking for excitations with a
length scale much larger than the lattice constant, we take the long-wave
approximation, that is, the continuum field theory approach will be employed
in the large $N$ limit. By assuming that the coherent amplitude is continuum
in space, the discrete variables $\alpha _{i}(t)$ can be replaced by a mean
field $\varphi\left(z,t\right)$, i.e. $\alpha _{i}\left(t\right) \rightarrow
\varphi \left( z,t\right)$. Correspondingly, the difference become the
following differential
\begin{equation}
\alpha _{i\pm 1}-\alpha _{i}\rightarrow \pm \frac{\partial }{\partial z}%
\varphi +\frac{1}{2}\frac{\partial ^{2}}{\partial z^{2}}\varphi +\cdots ,
\end{equation}%
where the lattice constant is assumed to be unity. By keeping the derivation
terms to the second order $\partial ^{2}/\partial z^{2}$, the system of
equations (\ref{coherent amplitude-Heisenbergeq}) become a field equation
with self-coupling terms
\begin{equation}
i\hbar \frac{d}{dt}\varphi +c_{0}S\frac{\partial ^{2}}{\partial x^{2}}%
\varphi +2c_{1}\varphi \left\vert \varphi \right\vert ^{2}=V \varphi,
\label{nls-1}
\end{equation}
which is a typical NLS equation in a constant potential $V=2Sc_{1}+B$.

To separate the rapid oscillation in the inhomogeneous nonlinear equation~(%
\ref{nls-1}), we rewrite $\varphi\left(z,t\right)$ as
\begin{equation}
\varphi \left( z,t\right) =e^{-i\chi t}\phi\left(\xi,t\right) ,  \label{leta}
\end{equation}%
where $\xi =x\sqrt{\hbar /(c_{0}S)}$, $\chi =\left( 2Sc_{1}+B\right) /\hbar$%
, and $\phi\left(\xi,t\right)$ is the slow varying envelope. Then the
standard form of the NLS equation is obtained
\begin{equation}
i\phi _{t}+\phi _{\xi \xi }+2\frac{c_{1}}{\hbar }\left\vert \phi \right\vert
^{2}\phi =0,  \label{nls}
\end{equation}%
which has the soliton solution. Here, the strength of nonlinear term
\begin{equation}
c_{1}=\frac{\delta }{2}J\left( 3\cos ^{2}\theta -1\right)  \label{c1}
\end{equation}%
is determined by spin-spin interaction parameter $J$ and the angle $\theta $
of the oblique magnetic field. We notice that when $\theta =\theta
_{0}=\arccos \sqrt{1/3}$, the nonlinear term in the equation of the motion
disappears, therefore Eq.(\ref{nls}) becomes a standard wave equation, which
gives a complete set of plane waves solution. In this case, the collective
excitations are spin waves. $\theta _{0}$ is named as the magic angle. It
shows that the magic angle changes the anisotropic system to the isotropic
one.

\section{\label{Sec:time-e}nonlinear time evolution of localized
magnetization}

For this ferromagnetic spin chain, the interaction parameter $J$ is
positive. Thus the sign of coefficient $c_{1}$ is determined by the angle of
the oblique magnetic field $\theta $. And nonlinear property of this system
is strongly related with the sign of the nonlinear term. With these
different physically accessible parameters, the nonlinear equation (\ref{nls}%
) can possess the bright-soliton and the dark-soliton solution. Next, we use
the inverse scattering method \cite{Dodd-b82,Ablowitz-b81} to obtain the
different solitons of this system.

When $c_{1}>0$, that is $\delta \left( 3\cos ^{2}\theta -1\right) >0$, the
single bright soliton solution of Eq.(\ref{nls-1}) is obtained as
\begin{equation}
\varphi \left( x,t\right) =Ae^{i\left( \gamma x-\omega t\right) }\text{sech}%
\left[ A\sqrt{\frac{c_{1}}{c_{0}S}}\left( x-x_{0}-vt\right) \right] ,
\label{b-soliton}
\end{equation}
with $v=v_{1}\sqrt{c_{0}S/\hbar },\gamma =\hbar v/2c_{0}S$ and
\begin{equation}
\omega =\frac{ c_{1}}{\hbar }\left( 2S-A^{2}\right) +\frac{B}{\hbar }+\frac{%
\hbar v^{2}}{4c_{0}S}
\end{equation}
where $v_{1}$ is an integral constant. The coefficient $v=v_{1}\sqrt{%
c_{0}S/\hbar }$ describes the velocity of bright soliton traveling to the
right. The positive coefficient $A$ characterizes the size of the bright
soliton. The coefficient $x_{0}$ denotes the initial position of the bright
soliton. Parameters $v_{1}$, $A$ and $x_{0}$ are determined by the initial
state. The bright soliton solution~(\ref{b-soliton}) depicts a wave packet
traveling in the continuous background.

When $c_{1}<0$, that is $\delta \left( 3\cos ^{2}\theta -1\right) <0$, we
can get a single dark-soliton solution of Eq.(\ref{nls-1})%
\begin{equation}
\varphi \left( x,t\right) =A^{\prime }e^{i\left( \gamma ^{\prime }x-\omega
^{\prime }t\right) }\tanh \left[ A^{\prime }\sqrt{\frac{-c_{1}}{c_{0}S}}%
\left( x-x_{0}^{\prime }-v^{\prime }t\right) \right]  \label{d-soliton}
\end{equation}
with $v^{\prime }=v_{1}^{\prime }\sqrt{c_{0}S/\hbar },$ $\gamma ^{\prime }=%
\frac{\hbar v^{\prime }}{2c_{0}S},$ and
\begin{equation}
\omega ^{\prime }=2\frac{c_{1}}{\hbar }\left( S-A^{\prime 2}\right) +\frac{B%
}{\hbar }+\frac{\hbar v^{\prime 2}}{4c_{0}S}
\end{equation}
where $v_{1}^{\prime }$ is an integral constant. $v^{\prime }=v_{1}^{\prime }%
\sqrt{c_{0}S/\hbar }$ characterizes the velocity of dark soliton traveling
to the right. The positive coefficient $A^{\prime }$ characterizes the size
of the dark soliton. The coefficient $x_{0}^{\prime}$ denotes the initial
position of the dark soliton. Parameters $v^{\prime}$, $A^{\prime}$ and $%
x_{0}^{\prime}$ are decided by the initial state. The dark soliton solution
describes a localized dip in the continuous background. Fig.~\ref%
{fig:standard} numerically illustrates the bright and the dark soliton
obtained from Eq.~(\ref{b-soliton}) and~(\ref{d-soliton}) respectively in
comparison with the results with many extra nonlinear terms in the following
discussion.
\begin{figure}[tbph]
\includegraphics[width=8 cm]{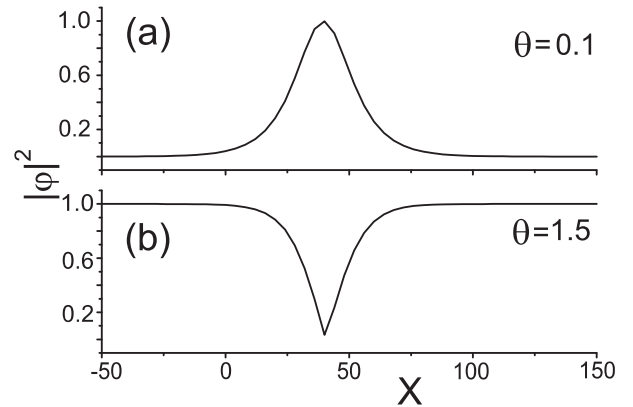}
\caption{Bright soltion (a) obtained from Eq.~(\protect\ref{b-soliton}) for $%
\protect\theta=0.1$. Dark soltion (b) obtained from Eq.~(\protect\ref%
{d-soliton}) for $\protect\theta=1.5$. Other parameters are taken as follows
$S=10, B=100, J=1, \protect\delta =0.1, x_{0}=x_{0}^{\prime }=0, A=A^{\prime
}=1, v_{1}=v_{1}^{\prime }=5$ and $t=3$. }
\label{fig:standard}
\end{figure}

Furthermore, to test the rationality for neglecting the terms $H_{\pm 1}$
and $H_{\pm 2}$, we deal with Hamiltonian (\ref{h-comp}) by the approach
presented above. The Holstein-Primakoff transformation (\ref{hp}) is first
employed to write Hamiltonian (\ref{h-comp}) in terms of bosonic operators
\begin{widetext}
\begin{eqnarray}
H &=&-c_{0}S\sum_{j}\left( \hat{a}_{j}^{\dag }\hat{a}_{j+1}+\hat{a}_{j}\hat{a%
}_{j+1}^{\dag }\right) +B\sum_{j}n_{j} -\left(
c_{0}+c_{1}\right) \sum_{j}n_{j}\left( n_{j+1}-2S\right)  \notag \\
&&+\frac{c_{0}}{4}\sum_{j}\left( \hat{a}_{j}^{\dag }n_{j}\hat{a}_{j+1}
+\hat{a}_{j}^{\dag }n_{j+1}\hat{%
a}_{j+1}+h.c.\right)+c_{2}\sqrt{2S}\sum_{j}\left( \hat{a}_{j}^{\dag }
n_{j\pm 1}+\frac{1}{2}\hat{a}_{j}^{\dag }n_{j}-2S\hat{a}_{j}^{\dag }+h.c.\right)  \notag \\
&&-2c_{3}S\sum_{j}\left( -\frac{1}{4S}\left( \hat{a}_{j}^{\dag }
n_{j}\hat{a}_{j+1}^{\dag }
+\hat{a}_{j}^{\dag }\hat{a}_{j+1}^{\dag }n_{j+1}\right) +\hat{a}_{j}^{\dag }%
\hat{a}_{j+1}^{\dag }+h.c.\right),  \label{h-comp-hp}
\end{eqnarray}
\end{widetext}
where $c_{0}=J\left( 1+\delta \sin^{2}\theta /2 \right)$, $c_{1}=J\delta
\left( 3\cos ^{2}\theta -1\right)/2$, $c_{2}=J\delta \sin 2\theta /4$ and $%
c_{3}=J\delta \sin ^{2}\theta/4$. Hamiltonian (\ref{h-comp-hp})
characterizes the low energy nonlinear property of ferromagnetic spin chain
in an oblique magnetic field. The corresponding Heisenberg equation
\begin{widetext}
\begin{eqnarray}
i\hbar \frac{d\hat{a}_{j}}{dt} &=&-c_{0}S \hat{a}_{j\pm1}
+\left( B+2S\left( c_{0}+c_{1}\right) \right) \hat{a}_{j}
-\left( c_{0}+c_{1}\right) \hat{a}_{j}n_{j\pm1}
+\frac{c_{0}}{4}\left( 2n_{j}\hat{a}_{j\pm 1}+%
n_{j\pm 1}\hat{a}_{j\pm 1}+\hat{a}_{j\pm
1}^{\dag }\hat{a}_{j}\hat{a}_{j}\right)  \notag \\
&&+c_{2}\sqrt{2S}\left( n_{j\pm 1}+\hat{a}_{j\pm 1}^{\dag }\hat{a}_{j}+\hat{a}_{j\pm
1}\hat{a}_{j}-2S +n_{j}+\frac{1}{2}\hat{a}_{j}\hat{a%
}_{j}\right)  \notag \\
&& -c_{3}2S\left( \hat{a}_{j\pm 1}^{\dag }-\frac{1}{2S}n_{j}\hat{a}_{j\pm 1}^{\dag }
-\frac{1}{4S}\hat{a}_{j\pm 1}^{\dag }n_{j\pm 1}
-\frac{1}{4S}\hat{a}_{j}\hat{a}_{j}\hat{a}_{j\pm 1}\right)
\label{comp-operator-Heisenbergeq}
\end{eqnarray}
\end{widetext}
contains rich nonlinear couplings and thus can predict various nonlinear
phenomena.

Further we use the continuum field theory approach, that is, expressing the
spin dynamics according to its corresponding continuous c-number equation in
Glauber's coherent-state representation. Then a similar field equation is
obtained as
\begin{eqnarray}
&&i\hbar \frac{d}{dt}\varphi +c_{0}S\frac{\partial ^{2}}{\partial x^{2}}%
\varphi +2c_{1}\varphi \left\vert \varphi \right\vert ^{2}  \notag \\
&=&\left( 2Sc_{1}+B\right) \varphi +2c_{2}\sqrt{2S}\left( \frac{5}{2}%
\left\vert \varphi \right\vert ^{2}+\frac{5}{4}\varphi ^{2}-S\right)  \notag
\\
&&-c_{3}\left( 4S\varphi ^{\ast }-3\varphi ^{\ast }\left\vert \varphi
\right\vert ^{2}-\varphi ^{3}\right) .  \label{comp-nls-1}
\end{eqnarray}
Obviously, the effects on $H_{\pm 1}$ and $H_{\pm 2}$ is kept in the above
derivation. Eq.(\ref{comp-nls-1}) is similar to the NLS equation, but more
nonlinear terms are involved, therefore the analytic solution can not be
got.
\begin{figure}[tbph]
\includegraphics[width=8 cm]{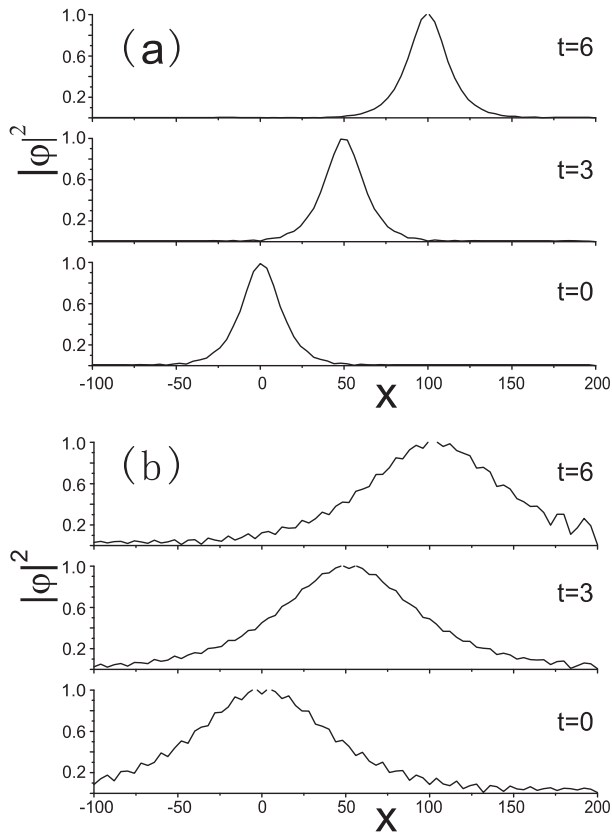}
\caption{Numerical time evolution of a bright soliton under the act
of Eq.~(\protect\ref{comp-nls-1}) for $\theta=0.1$ (a) and
$\theta=0.9$ (b). Other parameters are taken as follows $S=10,
B=100, J=1$ and $\delta =0.1$. It indicates that a bright soliton is
excited when $\theta<\theta_{0}$.} \label{fig:bright}
\end{figure}

\begin{figure}[tbph]
\includegraphics[width=8 cm]{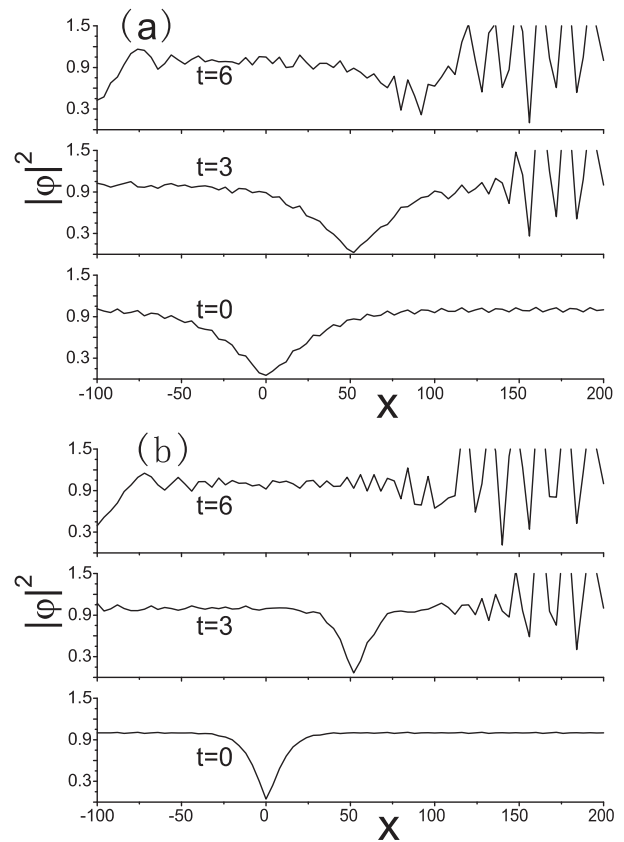}
\caption{Numerical time evolution of a dark soliton by the influence
of extra nonlinear terms in Eq.~(\ref{comp-nls-1}) for $\theta =1.0$
(a) and $\theta =1.5$ (b). Other parameters are taken as follows
$S=10,B=100,J=1$ and $\delta =0.1$. It indicates that a dark soliton
is excited when $\theta $ takes value on the right side of the magic
angle.} \label{fig:dark}
\end{figure}

To display the effects of extra nonlinear terms, we numerically
investigate the time evolution of dark and bright solitons under the
action of differential equation~(\ref{comp-nls-1}) in
Fig.~\ref{fig:bright} and Fig.~\ref{fig:dark}.
Figures~\ref{fig:bright} and~\ref{fig:dark} show that a bright and a
dark solition can be excited when $\theta$ takes value on the left
and the right side of the magic angle respectively. Comparing the
numerical results shown in Fig.~\ref{fig:bright} and~\ref{fig:dark},
it is found that the bright soliton is more stable than the dark
soliton solution. This is because the long time evolution of the
dark soliton in Fig.~\ref{fig:dark} become a profile with spatial
oscillation. While the bright soliton will almost keep their shapes,
the dark soliton will disappear after a long enough evolution.
Thus the bright soliton is more easily detectable in practice. From Fig.~\ref%
{fig:bright}, we found that the closer to the magic angle the tilted angle
is, the wider the soliton becomes.

\begin{figure}[tbph]
\includegraphics[width=3.2in]{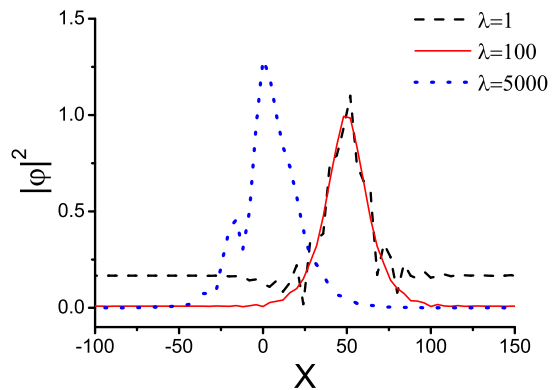}
\caption{(Color online) the time evolution of a bright soliton under
the effluence of extra nonlinear terms in Eq.~(\ref{comp-nls-1})
with respect to different values of the dimensionless parameter
$\lambda=B/J$ at $\theta=0.1, S=10, \delta =0.1, t=3$.} \label{blt3}
\end{figure}

In Fig.~\ref{blt3}, we plot the time evolution for the stable bright
soliton with respect to different values of the dimensionless
parameter $\lambda=B/J$. It can be found that the extra nonlinear
terms has no effect on solitons when $\lambda=100$. Such numerical
analysis confirms that the soliton-like wave can be excited under an
appropriate regime from $\lambda=10 $ to $1000$ approximately.
Therefore in this regime, it is reasonable to discard the terms
$H_{\pm1}$ and $H_{\pm2}$ in Eq.~(\ref{h-comp}). When parameters
$\lambda$ goes to infinite (see the line of $\lambda=5000$ in
Fig.\ref{blt3}), the interaction term $H_{0}$ can be completely
ignored, then Hamiltonian of the system approximately described by
$BL_{z}$. In this situation, the system is completely polarized
along the direction of the magnetic field and form a background with
symmetry breaking which is displayed by the dot line in Fig.~
\ref{blt3}. The numerical result that a very sharp peak is localized
around $x=0$ justifies our observation from the physical intuition.

There are some experimental data on quasi-one-dimensional
ferromagnetic chains
\cite{Jain-prb06,Rubins-pra00,Greeney-prb89,Montfrooij-prb01,Torrance-pr69},
which may be used to test our predictions in an indirect way. For
the spin chain made of the material
Ca$_{3}$Co$_{2-x}$Fe$_{x}$O$_{6}$ (x=0, 0.1, 0.2) (Ref.
\cite{Jain-prb06}), the exchange constant $J$ decreases from $4.39$
to $8.13K$ and the magnetic field strength $B<10T$ ($\sim 33.6 K$).
The solitary-excitation is possible since the dimensionless
parameter $\lambda=B/J\approx10$. However, for the three
ferromagnetic
chains\cite{Rubins-pra00,Greeney-prb89,Montfrooij-prb01,Torrance-pr69},
which made of the following material: (1)
[(CH$_{3}$)$_{3}$NH]FeCl$_{3}\cdot$2H$_{2}$O (Ref.
\cite{Rubins-pra00,Greeney-prb89}), where $J\sim 17.4K, B<16T (\sim
21.5 K)$; (2) CoCl$_{2}\cdot$2D$_{2}$O(Ref.
\cite{Montfrooij-prb01}), where $J\sim 2.475K$ and  $B<6T$ ($\sim
12.1 K$); (3) CoCl$_{2}\cdot$H$_{2}$O(Ref. \cite{Torrance-pr69}),
$J\sim 18.3K$ and $B<6T$ ($\sim 12.1 K$), the bright soliton may be
excited, but noises exist as shown in Fig.~\ref{blt3}, with the
dimensionless parameter $\lambda=B/J\approx1$.

\section{\label{Sec:conclusion} Remarks and conclusion}

Before concluding this paper, let us discuss the physical meaning of the
solitary wave of magnetic excitation in a $N$-site $XXZ$ quantum spin chain
from the point view of quantum information processing. We first note that
the multi-mode coherent states $\left\vert\alpha \right\rangle
=\prod_{j}\left\vert \alpha _{j}\right\rangle$ represent an inhomogeneous
collective excitations distributed around the spin chain. This collective
excitation has a spin representation by
\begin{equation}
\hat{a}_{j}^{\dag }=L_{i}^{+}\frac{1}{\sqrt{S-L_{i}^{z}}}\equiv QL_{i}^{+}
\end{equation}%
with $Q=1/\sqrt{S-L_{i}^{z}+1}$ and $\left\vert 0\right\rangle
_{j}=\left\vert S,-S\right\rangle _{j}$ being the lowest eigenstate of the
on-site spin $\mathbf{L}_{j}$. We also notice that the coherent state
denotes a superposition of various spin states $\left\vert
S,m_{s}\right\rangle _{j}$, i.e.,
\begin{equation}
\left\vert \alpha _{i}\right\rangle =\sum C_{n}(\alpha
_{i})(QL_{i}^{+})^{n}\left\vert S,-S\right\rangle _{j}=\sum B_{n}(\alpha
_{i})\left\vert S,n-S\right\rangle _{j}
\end{equation}%
where $C_{n}(\alpha _{i})=\exp (-|\alpha _{j}|^{2}/2)\alpha
_{j}^{n_{j}}/(n_{j}!)$ and $B_{n}(\alpha _{i})$ can be obtained from
$C_{n}(\alpha _{i})$ directly. During the propagation of soliton,
the narrow traveling wave packet does not spread, therefore it
behaves like a ``flying qudit'' (a $d=2S$ level system).

When $S=1/2,$ the qubit is localized at the $j$th site with a superposition
state
\begin{equation}
\left\vert\alpha _{j}\right\rangle\sim\left\vert \frac{1}{2},-\frac{1}{2 }%
\right\rangle+\frac{\alpha _{j}}{2}\left\vert \frac{1}{2},\frac{1}{2}%
\right\rangle
\end{equation}
which can be used to encode quantum information as usual. During the
propagation, the wave function nearly keeps its spatial shape all the time.
From an mathematical point of view, the spatially non-spreading properties
of the carrying excitation wave is very crucial for quantum state transfer
from one location to another with high fidelity. It seems the bosonic
excitations obey the bosonic commutation relations only in the large $S$
limit, but the spin wave approach can still work well for $S=1/2$ in the
condensed matter physics. In this sense we can suppose our above arguments
available.

In summary, we have studied solitary magnetic excitation in a
$N$-site $XXZ$ quantum spin chain as well as how to use an oblique
magnetic field to create different types of solitons. Through a mean
field approximation beyond the usual spin wave approach, we obtain
the quasi-classical motion equations for nonlinear evolution of the
Heisenberg spin system. We show that the switch between the bright
and the dark soliton is controlled by the angle of the magnetic
field, whose tangent is defined by the ratio between the z component
and the x component of the magnetic field. And at the magic angle,
the system go to the isotropic Heisenberg model, hence, only an
ideal spin wave is stimulated. We also remark the possibility for
solitons to play the role of ``flying qudit'' based on the
well-known results in Ref.~\cite{yangshuo} that a no-spreading
wave-packet behaves like a ``flying qubit''.

This work is supported by the NSFC with grant Nos. 90203018,
10474104, 60433050, 10325523, 10347128, 10075018 and 10704023, the
NFRPC with Nos. 2001CB309310 and 2005CB724508, and the Scientific
Research Fund of Hunan Provincial Education Department of China
(Grant No. 07C579).

\end{document}